\documentclass[aps,prc,twocolumn,showpacs,groupedaddress,floatfix,10pt]{revtex4-1}
\usepackage{dcolumn}
\usepackage{bm}
\usepackage{amsfonts}

\usepackage{bbold}

\usepackage{amssymb, amsmath}
\usepackage{graphicx}

\usepackage{xspace}
\usepackage{xcolor}
\usepackage{verbatim}

\newcommand{\veps}{\varepsilon}

\newcommand{\unedf}{\texttt{UNEDF1}\xspace}
\newcommand{\slyd}{\texttt{Sly4d}\xspace}

\begin{document}
\title{Dynamical effects in fusion with exotic nuclei}

\author{K. Vo-Phuoc}
\email{kirsten.vo-phuoc@anu.edu.au}
\affiliation{Department of Nuclear Physics, Research School of Physics and Engineering, The Australian National University, Canberra ACT  2601, Australia}

\author{C. Simenel}
\affiliation{Department of Nuclear Physics, Research School of Physics and Engineering, The Australian National University, Canberra ACT  2601, Australia}

\author{E. C. Simpson}
\affiliation{Department of Nuclear Physics, Research School of Physics and Engineering, The Australian National University, Canberra ACT  2601, Australia}

\date{\today}

\begin{abstract}
\begin{description}
\item[Background] Reactions with stable beams have demonstrated a
  strong interplay between nuclear structure and fusion.  Exotic beam
  facilities open new perspectives to understand the impact of neutron
  skin, large isospin, and weak binding energies on fusion.
  Microscopic theories of fusion are required to guide future
  experiments.
\item[Purpose] To investigate new effects of exotic structures and
  dynamics in near-barrier fusion with exotic nuclei.
\item[Method] Microscopic approaches based on the Hartree-Fock (HF)
  mean-field theory are used for studying fusion barriers in
  $^{40-54}$Ca+$^{116}$Sn reactions for even isotopes.  Bare potential
  barriers are obtained assuming frozen HF ground-state densities.
  Dynamical effects on the barrier are accounted for in time-dependent
  Hartree-Fock (TDHF) calculations of the collisions.  Vibrational
  couplings are studied in the coupled-channel framework and
  near-barrier nucleon transfer is investigated with TDHF
  calculations.
\item[Results] The development of a neutron skin in exotic calcium
  isotopes strongly lowers the bare potential barrier. However, this
  static effect is not apparent when dynamical effects are
  included. On the contrary, a fusion hindrance is observed in TDHF
  calculations with the most neutron rich calcium isotopes which
  cannot be explained by vibrational couplings. Transfer reactions are
  also important in these systems due to charge equilibration
  processes.
\item[Conclusions] Despite its impact on the bare potential, the
  neutron skin is not seen as playing an important role in the fusion
  dynamics.  However, the charge transfer with exotic projectiles
  could lead to an increase of the Coulomb repulsion between the
  fragments, suppressing fusion. The effect of transfer and
  dissipative mechanisms on fusion with exotic nuclei deserve further
  studies.
\end{description}
\end{abstract}

\maketitle

\section{Introduction}

Heavy-ion collision studies rely on a good understanding of the interplay
between the structure of the collision partners and reaction mechanisms.  This
is particularly important at near barrier energies, where complex quantum
effects such as tunnelling and coherent coupling between reaction channels are
magnified.  In particular, these quantum effects have a strong impact on fusion
between two nuclei and are highly sensitive to the structure of these nuclei.
For instance, a variation of few neutrons in the choice of the target could lead
to variations of sub-barrier fusion cross-sections by orders of magnitudes
\cite{rei85,lei95}.

In fact, the discovery that fusion is strongly influenced by the initial
structure of the reactants came as a surprise \cite{sto76,Beckerman80}.  Indeed,
the collision partners quickly lose their identity when they merge, on typical
time scales of few zeptoseconds ($10^{-21}$~s).  Nevertheless, this time frame is
sufficiently long to enable
couplings between the relative motion and internal collective excitations
\cite{esb81,das83a}.  These couplings lead to a structure dependent distribution
of fusion barriers \cite{row91}, a phenomenon which has been widely studied
experimentally (see Refs. \cite{das98,bac14} for reviews) and traditionally
interpreted within the coupled-channel framework (see Refs.
\cite{bal98,das98,hag12,bac14} for reviews).  Coupling to rotational
\cite{won73,sto78,lei95} and low-lying vibrational
\cite{esb81,rei85,Morton94,lei95,ste95a} states, as well as to transfer
\cite{Beckerman80,Stefanini86,Timmers97,bou14,jia14,Jiang15} and breakup
of colliding partners 
\cite{Dasgupta99,can06,kee07} have been shown to have a strong
impact on fusion.

This modern picture of heavy-ion fusion has been achieved thanks to
high-precision measurements with stable beams.  The recent development
of exotic beam facilities has now opened new perspectives to fusion
studies.  In particular, the role of large neutron halos or skins,
soft dipole resonances, weak nucleon binding energy, and large isospin
asymmetry could be systematically investigated experimentally in the
near future. All are expected to impact fusion
\cite{tak91,hus91,das92}.  The first fusion studies with exotic beams
focussed on reactions with light neutron-rich nuclei to understand the
impact of their neutron excess and enhanced breakup and transfer due
to weak neutron binding energies
\cite{zyr97,you98,tro00,raa04,Lemasson11}.  More recent experiments
have used heavier exotic beams, such as $^{132}$Sn and $^{134}$Te, to
study the interplay between transfer and fusion \cite{koh11,koh13}.

The purpose of this work is to study the fusion mechanism away from
stability.  We focus on the fusion barrier, $V_B$, which is sensitive to
the structure of the collision partners.  Fusion barriers were
systematically studied for stable nuclei as soon as heavy-ion beams
with sufficient energies were available \cite{vaz81,new04}.  The
barrier is generated by the competition between Coulomb repulsion and
strong nuclear attraction between the fragments.  
Exotic structures, for example large neutron skins, could affect the
barrier radius, $R_B$, and in turn $V_B$.  A legitimate question is also to ask how these
effects of nuclear structure on the fusion barrier would be impacted
by the reaction dynamics \cite{agu92}.


In order to investigate both static and dynamic effects on fusion with exotic
nuclei, we use a microscopic approach based on the Hartree-Fock (HF) mean-field
approximation.  In this approximation, each nucleon is assumed to move
independently in the self-consistent mean-field generated by the other nucleons.
Static HF calculations account for important nuclear structure characteristics
such as shell effects \cite{vau72}, ground-state deformation \cite{vau73} and
neutron skin \cite{rot09}.  Dynamical effects can also be accounted for in
time-dependent Hartree-Fock (TDHF) theory which allows the nuclear density and
thus the mean-field to evolve in time.  The fact that TDHF calculations treat
static and dynamical effects on the same footing is particularly attractive to
study the interplay between nuclear structure and reaction mechanisms (see Refs.
\cite{neg82,sim12b} for reviews).  Another invaluable feature for exotic systems
is that no prior knowledge of the structure of the reactants is required.
Indeed, the only parameters are those of the effective interaction between
nucleons, usually of the Skyrme type \cite{sky56}.  For instance, time-dependent
microscopic calculations have been successful in reproducing fusion reactions
with exotic $^{132}$Sn beam \cite{obe13}.


We focus our study on collisions of calcium isotopes with $^{116}$Sn.  The
static effect of neutron skin on the bare potential barrier is first studied in
section~\ref{sec:static} with the frozen-HF method.  TDHF calculations are then
presented in section~\ref{sec:tdhf} to investigate the net effect of the
dynamics on the fusion thresholds for these systems.  In order to understand the
contribution of the vibrational couplings to the dynamics we perform
coupled-channel calculations in section~\ref{sec:cc} where the properties of the
vibrational states are calculated with a TDHF code using linear response
theory.  Finally, the importance of transfer channels is investigated with TDHF
calculations in section~\ref{sec:transfer}.

\section{Static effects}\label{sec:static}
\subsection{The frozen Hartree--Fock approach}

The nucleus-nucleus potential between two ground state nuclei, otherwise known
as the bare potential where the densities of the nuclei stay ``frozen" at all
points \cite{bru68}, was calculated using the frozen HF method
\cite{den02,sim13c}.  One first has to separately compute the HF ground-states
of the nuclei.  Then, the total energy of the system $E(\mathbf{R})$ is computed
from the total density by setting a distance vector $\mathbf{R}$ between the
centres of mass.

The nuclear part of the nucleus-nucleus potential, 
denoted by $V$, is given by \cite{bru68},
\begin{equation}
\label{vint} V(\mathbf{R}) = E(\mathbf{R})-E_{\text{HF}}[\rho_1]-E_{\text{HF}}[\rho_2],
\end{equation}
where $\mathbf{R}$ is the position vector between the centres of masses of the
two separate systems which have ground state densities $\rho_1$ and $\rho_2$
respectively. The total interaction energy can be written as an integral of a
local energy density function,
\begin{align*}
  E(\mathbf{R}) &= 
    \int \mathcal{E}\left[\rho_1(\mathbf{r})+\rho_2(\mathbf{R-r})\right] 
    d\mathbf{r}.
\end{align*}
The HF energy for each nucleus ($j = 1,2$) reads
\begin{align*}
  E_{\text{HF}}[\rho_j] &= \int \mathcal{E}[\rho_j(\mathbf{r})] d\mathbf{r}.
\end{align*}
The same Skyrme energy density functional \cite{sky56} is used to
compute both ground-state densities and the nuclear interaction
between the nuclei.  Then, the set of parameters of the Skyrme
functional is the only input needed to compute the frozen HF
potential.  Two parametrisations of the Skyrme interaction,
the \slyd \cite{kim97} and \unedf \cite{kor12} parametrisations, have been
used for the most part of this work.  Both ignore the centre of mass
corrections in the fitting procedure so they are suitable not just for
static calculations but also dynamical simulations \cite{sim12b,
  kor12}.

The HF ground-states of the nuclei were calculated using the \verb+ev8+
\cite{bon05} code.  Pairing correlations were included at the BCS level with a
surface pairing interaction \cite{ben03} with a density dependent pairing force
\cite{bender99,kri90} of the form
\begin{align}
  {v}_{\text{pair}}({\mathbf{r}}_1, {\mathbf{r}}_2)
    = \tilde{t_0}\,\delta\left({\mathbf{r}}_1-{\mathbf{r}}_2\right)
      \left(1-\frac{\rho(\tilde{\mathbf{r}})}{\tilde{\rho}_0} \right), \label{eq:pair}
\end{align}
where $\tilde{\mathbf{r}}=({\mathbf{r}}_1+{\mathbf{r}}_2)/2$ and with parameter
values $\tilde{t}_0 = 1000$ MeV fm$^3$ and $\tilde{\rho}_0 = 0.16$ fm$^{-3}$.
Pairing correlations are known to have a small effect on fusion \cite{sca14a}.
However, they avoid spurious ground-state deformations (by distributing the
occupation numbers near the Fermi surface) which could in turn have a spurious
effect on the prediction of the fusion barrier.

All nuclei studied here are spherical in their ground-state and were calculated
on a 3D grid with cubic box of size $22.4^3$ fm$^3$ with three planes
of symmetry and with mesh size $\Delta x=0.8$ fm.  All HF calculations
henceforth use this same mesh size.

The calculations for the frozen HF bare potential were done on a box size
$67.2\times22.4\times22.4$ fm$^3$ in the $x-y-z$ orientation where the
collision axis is the $x-$axis. The plane $z=0$ is a plane of symmetry.  The
maximum distance between the two centres of mass was 44.8~fm.

\begin{figure}[h]
  \centering
  \graphicspath{ {./pix/} }
  \includegraphics[width=8cm, clip=true, trim=0 5mm 0 2cm]{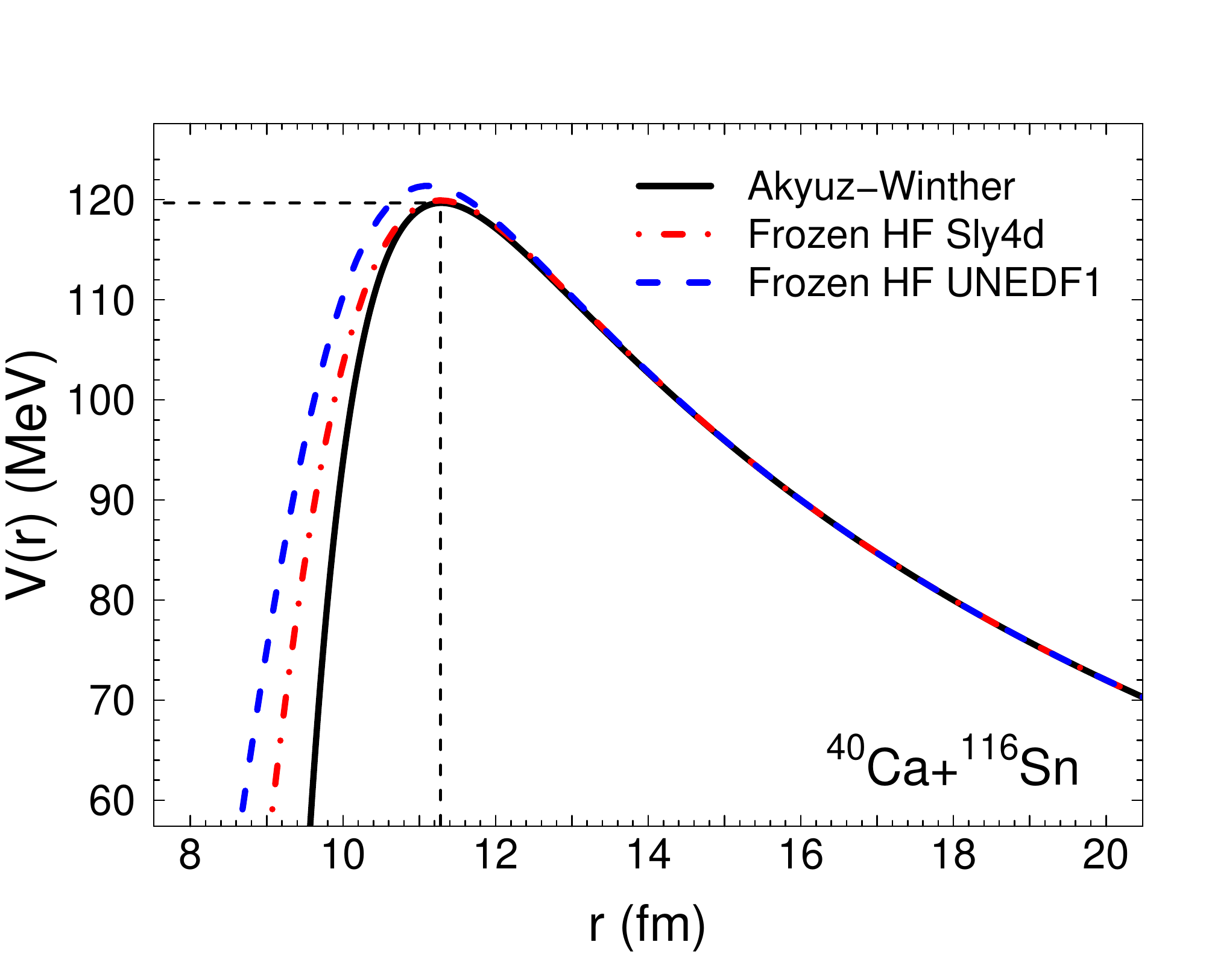}
  \caption{(Color online) Example of the bare nucleus-nucleus potential 
    from the frozen HF method (dashed and dot-dashed lines) and the Aky\"{u}z--Winther
    phenomenological potential \cite{BrogWinth} (solid line) for
    $^{40}$Ca+$^{116}$Sn.  The dashed lines show the maximum AW barrier energy
    at $V=119.7$ MeV and $r=11.3$ fm. }
  \label{fig:frozenpot}
\end{figure}

The resulting nucleus-nucleus potential obtained from the sum of the nuclear
part \eqref{vint} and the Coulomb potential is shown in Fig. \ref{fig:frozenpot}
for the sample system $^{40}$Ca+$^{116}$Sn.  The phenomenological
Aky\"{u}z--Winther (AW) \cite{BrogWinth} nucleus-nucleus potential is also shown
for comparison.

The agreement between the fully microscopic frozen HF approach and the
phenomenological potential in terms of height and position of the
barrier is remarkable.  Both approaches disagree, however, on the
inner part of the potential.  There is naturally a large uncertainty
in predicting the form of the potential in macroscopic
phenomenological approaches when the two nuclei strongly overlap
\cite{das03,esb06}.  
The frozen HF approach, however, does not rely on
an a priori guess of the form of the potential, and the microscopic
calculations predict a wider barrier than the AW potential.  Note that
the present frozen HF calculations naturally incorporate effects from
incompressibility \cite{esb06} but neglect the Pauli exclusion
principle between nucleons belonging to different nuclei.  The latter
is expected to reduce the attraction between the nuclei inside the
barrier and then to further increase the barrier width.  The inner
part of the potential affects deep sub-barrier fusion cross-sections
\cite{jia02,das07} but it is not expected to have a large impact on
near-barrier data from which the fusion barrier distributions
\cite{row91} are extracted experimentally \cite{lei95,das98}.
Therefore, a modification of the inner part of the potential would not
affect the conclusions of this work.

\subsection{Results}
\begin{figure}[h]
  \centering
  \graphicspath{ {./pix/} }
  \includegraphics[width=8cm, clip=true, trim=0 0 0 2cm]{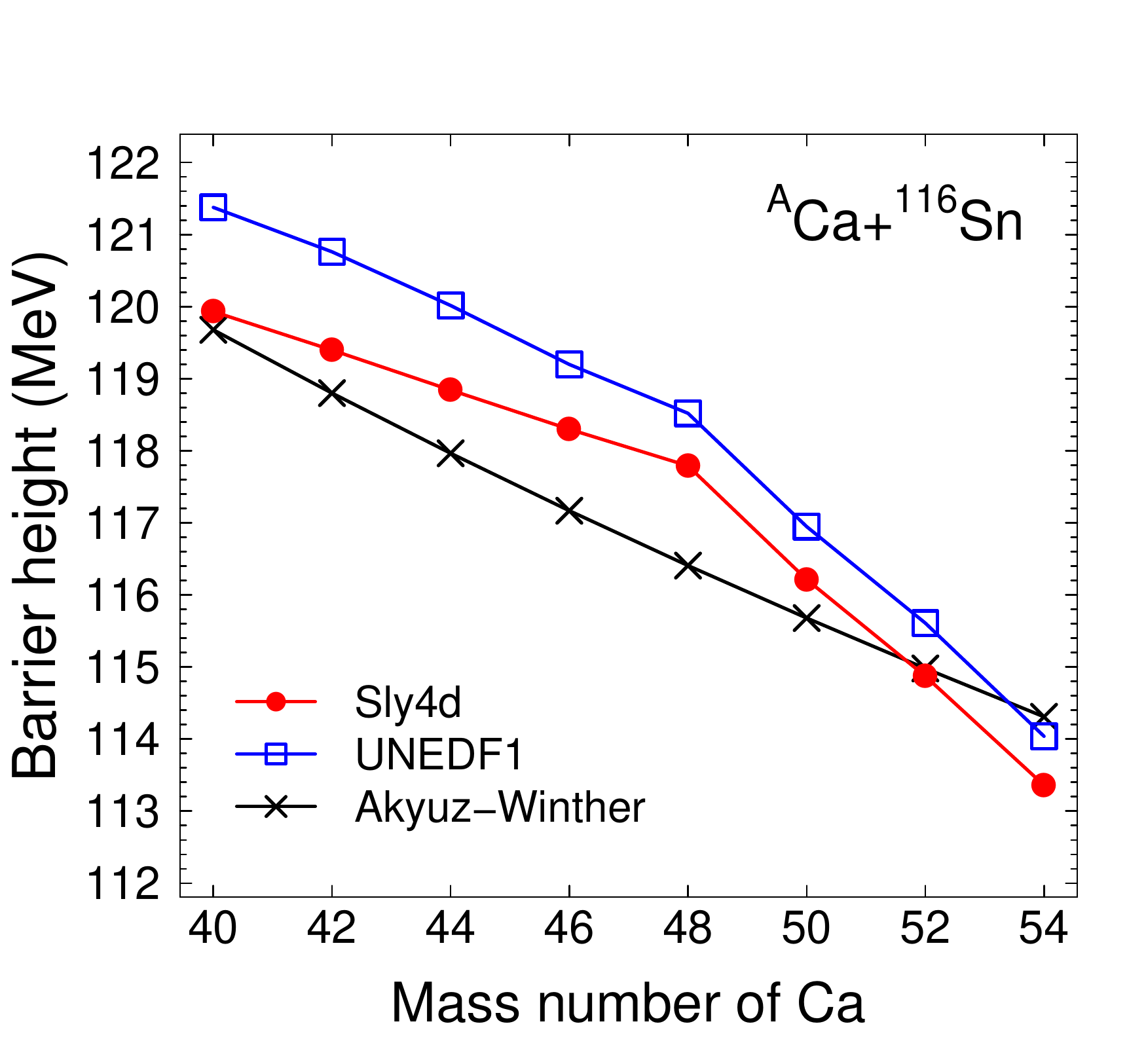}
  \caption{(Color online) Frozen HF barriers of calcium isotopes on $^{116}$Sn.
    Two different parametrisations are used. Also included
    are the macroscopic Aky\"{u}z--Winther potential barriers.}
  \label{fig:frozvb}
\end{figure}

The bare potentials in $^{A}$Ca$+^{116}$Sn systems have been computed with the
frozen HF approach for the \slyd and \unedf parametrisations of the Skyrme
interaction.  The resulting barrier energies, $V_B$, are presented in
Fig. \ref{fig:frozvb} together with fusion barriers obtained from the AW
potential.

\begin{figure}[h]
  \centering
  \graphicspath{ {./pix/} }
  \includegraphics[width=0.45\textwidth, page=3, clip=true, trim=0 0 0 2cm]{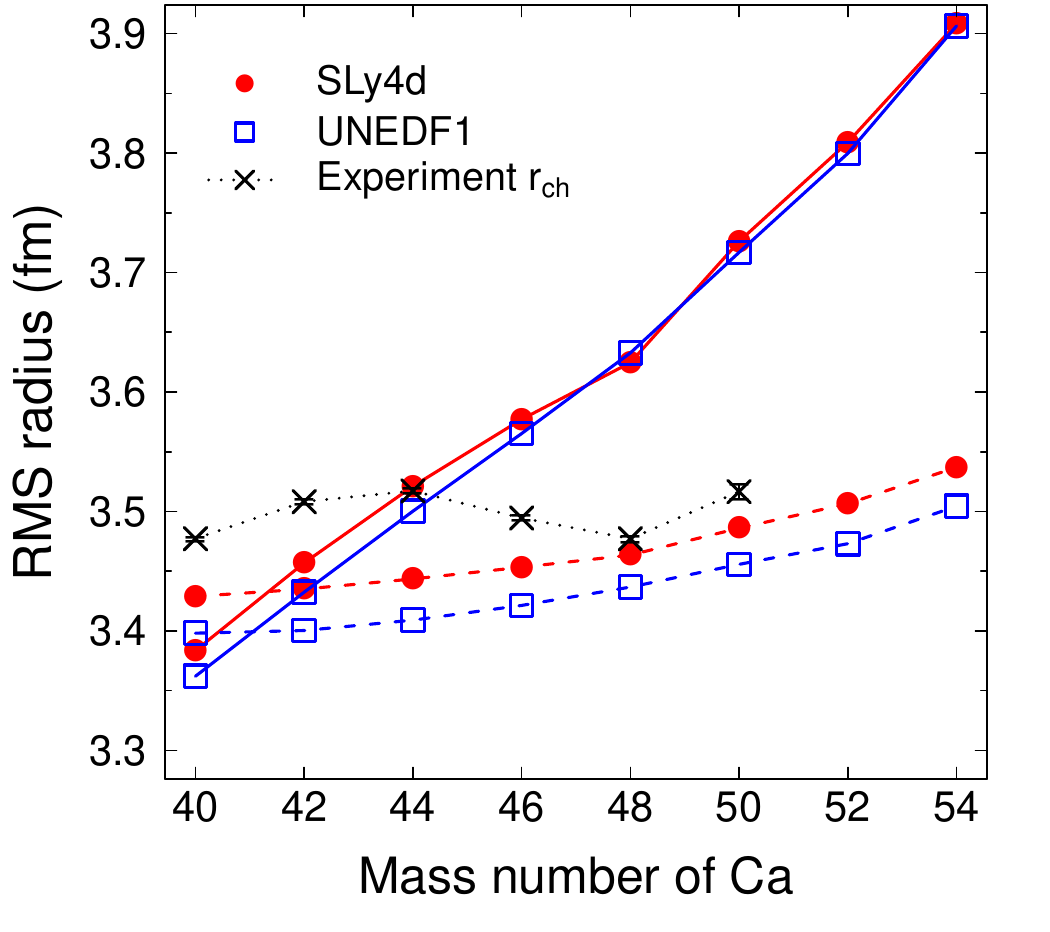}
  \caption{(Color online) HF proton (dashed lines) and neutron (solid lines) root mean square radii 
    in calcium isotopes for two parametrisations of the Skyrme functional.
    The experimental charge radii (crosses) are from Refs. \cite{woh81,ver92}.}
  \label{fig:radii}
\end{figure}

A reduction of the barrier height is observed with increasing $A$ in
each set of calculations due to the increase in size of the calcium
isotopes (see Fig.~\ref{fig:radii}).
However, the HF calculations for both Skyrme parameterizations also
exhibit a faster reduction of $V_B$ with increasing $A$ for $A\geq
48$.  This feature is not seen with the AW phenomenological potential,
which is a simple smooth function of the mass number $A$ and does not
account for quantum shell effects.


In order to interpret this change of behaviour near the doubly magic
$^{48}$Ca isotope in the microscopic calculations, the
root-mean-square (rms) radii, both proton and neutron, have been
computed for the calcium HF ground states.  These are shown in
Fig. \ref{fig:radii}.  There is a change of gradient in the neutron
rms radius at the $^{48}$Ca nucleus in both parametrisations,
indicating
the development of a neutron skin.  A more rapid increase of rms
radius after $^{48}$Ca means a more rapid decrease of barrier energies
due to the lowering of the Coulomb repulsion between the reactants.
Also included in Fig. \ref{fig:radii} are the experimental charge rms
radii \cite{woh81, ver92}. These values are relatively close to the
calculated proton radii, especially for the doubly magic nuclei.
The deviations are larger for mid-shell
nuclei due to correlations not accounted for at the mean-field level
\cite{cau01}.

The neutron single particle shell levels arising from the HF ground
states of the calcium nuclei help understand why the HF rms radius
behaves this way.  Between $^{40}$Ca to $^{48}$Ca the $1f_{7/2}$
neutron shell is progressively filled.
Then, after $^{48}$Ca there is a magic shell gap of approximately 5 MeV in
energy until the next $2p_{3/2}$ shell.
The last occupied shell in $^{54}$Ca is the $2p_{1/2}$ shell.
The $2p$ neutron levels are weakly bound in these calcium
isotopes and, in comparison to $1f$ levels, they also have an
additional node.  These effects can explain the faster increase of the
neutron radius with $A$ for isotopes heavier than $^{48}$Ca.


To conclude, the development of a neutron skin in exotic calcium
isotopes could significantly lower the barrier.  This phenomenon is
due to quantum shell effects and is not accounted for in standard
parametrisations of the nucleus-nucleus potential. The latter are then
only valid close to the valley of stability.  The lowering of the
barrier due to the neutron skin is purely a static effect and it remains
to investigate how this effect is affected by the dynamics.


\section{Dynamic effects}\label{sec:tdhf}
\subsection{The time-dependent Hartree--Fock approach}
Obtaining a barrier energy for fusion using frozen ground state densities is a
useful starting point but a more realistic barrier energy naturally should
include the dynamics of the nuclei as they approach each other,
for example the possibility of vibrations, rotations and transfer.
Traditionally this is done within the coupled-channel formalism. 
An alternative approach is to use the time-dependent Hartree-Fock
(TDHF) method.
Early TDHF applications already included vibration \cite{blo79}, fission
\cite{neg78}, and reaction \cite{bon76} studies.  However, realistic
calculations, including all dynamics at the mean-field level,
have only been made possible in the past decade thanks to the
development of three-dimensional TDHF codes
\cite{kim97,mar05,uma05,nak05,mar14}
including spin-orbit couplings \cite{uma86}.  In particular, the TDHF approach
has been shown to give different fusion thresholds than the corresponding
frozen HF bare potential computed with the same functional, indicating an
important role of dynamics on the fusion mechanism
\cite{sim08,was08,sim13b,sim13c,uma14}.

The TDHF equation reads
\begin{align*}
i\hbar \frac{d}{dt}\rho = \left[h[\rho],\rho\right],
\end{align*}
where $\rho$ is the one-body density matrix and $h[\rho]$ is the self-consistent
single-particle HF Hamiltonian.  TDHF codes solve this equation in the canonical
basis in which the elements of the one-body density matrix read
\begin{align*}
\rho(\mathbf{r},s,q;\mathbf{r}',s',q')=\sum_i n_i \, 
    \varphi_i(\mathbf{r},s,q)\, \varphi_i^*(\mathbf{r}',s',q'),
\end{align*}
where $\varphi_i$ are the single-particle wave-functions with occupation numbers
$n_i$, and $\mathbf{r},s,q$ denote position, spin and isospin, respectively.  In
this basis, the TDHF equation can be written as a set of non-linear
Schr\"odinger equations for each single-particle wave-function
\begin{align*}
i\hbar \frac{d}{dt}\varphi_i= h[\rho]\,\varphi_i.
\end{align*}
This set of equations is solved iteratively in time, with the HF Hamiltonian
determined at every iteration according to its relationship with the energy
density functional
\begin{align*}
h[\rho](\mathbf{r},s,q;\mathbf{r}',s',q')
    =\frac{\delta E[\rho]}{\delta\rho(\mathbf{r}',s',q';\mathbf{r},s,q)}.
\end{align*}

The  effects of dynamics on  fusion were studied here with the 
\verb+TDHF3D+ code \cite{kim97}.
The calculations start with the HF ground states of the nuclei put together in a
larger box as done for frozen HF calculations.  Now, a mean field for the entire
system is generated by all the independent particles from both nuclei.  A
Galilean boost is applied to each nucleus at initial time $t=0$ and the
mean-field evolves self-consistently \cite{bon76}.  The evolution of the
occupied single-particle wave-functions of the nuclei with respect to time is
computed as the nuclei move relative to each other.

The occupation numbers $n_i$ are those determined in the ground-state static
calculations by the pairing interaction~(\ref{eq:pair}) in the BCS
approximation.  These occupation numbers are kept constant during the dynamics,
that is, we used the so-called frozen occupation number approximation \cite{mar05}.
A more sophisticated approach would imply a BCS \cite{sca13b} treatment to allow
the occupation numbers $n_i$ to evolve in time.  Unlike in fission in which
dynamical pairing correlations play an important role \cite{neg78,sca15a,bul16},
they have been shown to barely affect fusion \cite{sca14a} and are neglected in
the present work.

The TDHF method is much more computationally demanding than its static
counterpart.  As for the frozen case, a plane of symmetry at $z=0$ is assumed in
the \verb+TDHF3D+ code to speed up numerical simulations.  All calculations were
done using the same box conditions as for the frozen HF calculations and with
both the \slyd and \unedf parametrisations of the Skyrme functional.  The
starting distance must be large enough so that Coulomb excitation is properly
accounted for in the approach phase.
This is particularly important for calculations involving large charge products
$Z_1Z_2$ \cite{wak14,uma15a}.  Starting at a separation distance between the centres 
of masses of 44.8 fm is large enough to
account for this effect in Ca$+$Sn collisions.  The time step size used between
each iteration was $1.5\times10^{-24}$~s to ensure convergence.  This study
focuses on $L=0$ angular momentum fusion barriers, therefore only central
collisions are considered.

To extract a fusion barrier energy from the TDHF method, the distance
between centre of masses of the two interacting nuclei was used.  The
notion of defining separate centres of mass in a single mean field
(for the whole system) is addressed by defining a neck plane
perpendicular to the collision axis.  At each time, its purpose is to
separate the entire density of the system into two fragments
\cite{was08}.  Using this, the centres of masses of these two
fragments and the distance between them can then be calculated at each
time step.  Fusion for the system was deemed successful if, at a given
centre of mass energy $E_{c.m.}$, the distance between the centres of
masses remained below $\sim10$ fm after $4.5$ zs.

\begin{figure}[h]
  \centering
  \graphicspath{ {./pix/} }
  \includegraphics[width=0.45\textwidth, clip=true, trim = 0 0 0 2cm]{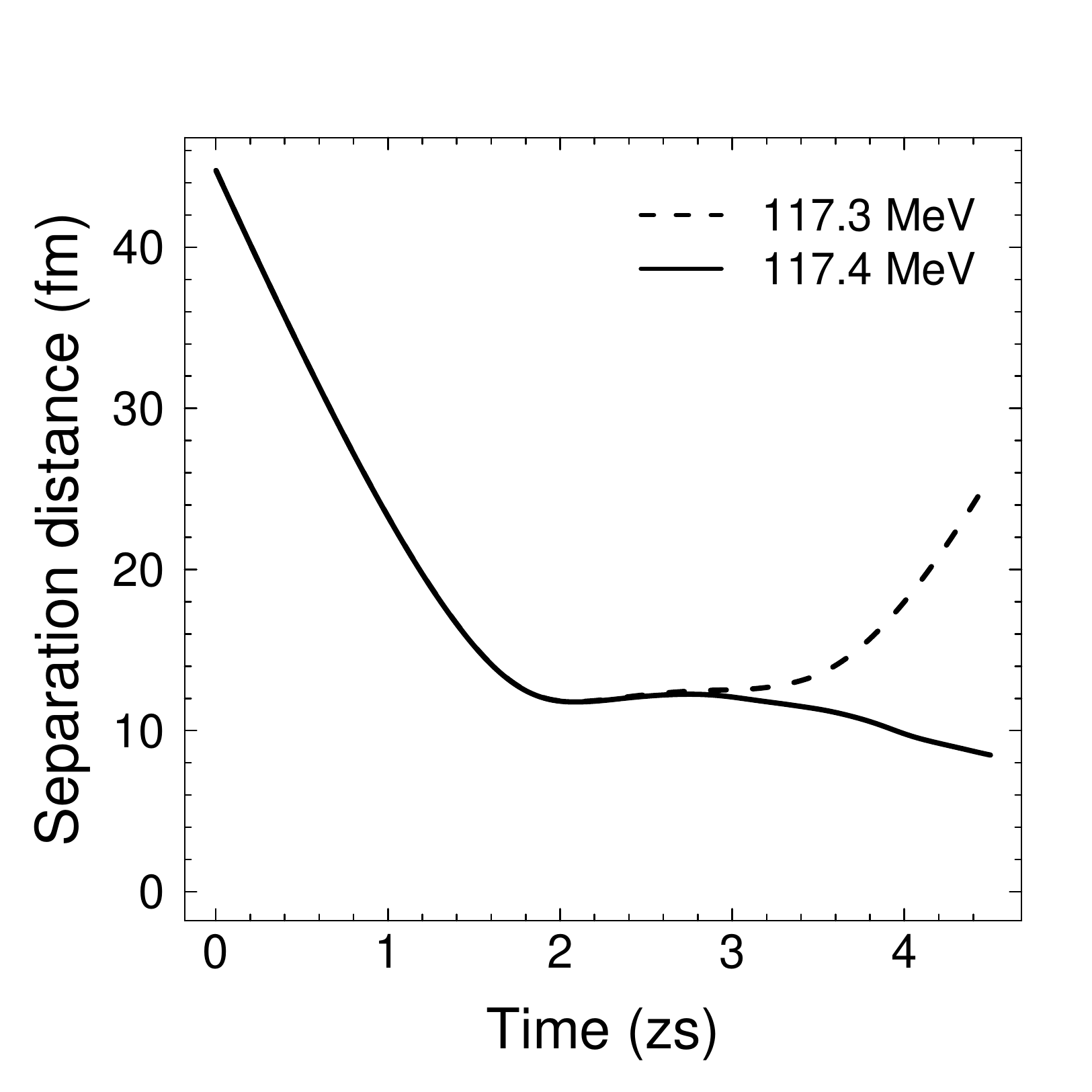}
  \caption{Distance between fragment centres of masses in $^{40}$Ca+$^{116}$Sn
    central collisions as a function of time. Fusion is achieved at $E_{c.m.}=
    117.4$ MeV (solid line) while separation of the nuclei occurs at
    $E_{c.m.}=117.3$~MeV (dashed line).}
  \label{fig:dist}
\end{figure}

An example of the distance between centre of mass versus time is given in
Fig. \ref{fig:dist} for $^{40}$Ca+$^{116}$Sn.  The solid line is associated with
a trajectory where fusion occurred.  The TDHF fusion probability for a given
initial condition is either 0 or 1 as there is no tunnelling of the many-body
wavefunction taken into account.  In TDHF calculations, the fusion barrier
energy is really then a fusion threshold.  This threshold is found by varying
$E_{c.m.}$ in steps of 0.1 MeV.  Each TDHF fusion threshold energy then has a
numerical uncertainty of $\pm0.05$~MeV.

\subsection{TDHF results}

\begin{figure}[!htb]
  \graphicspath{ {./pix/} }
\includegraphics*[width=8cm,page=4]     {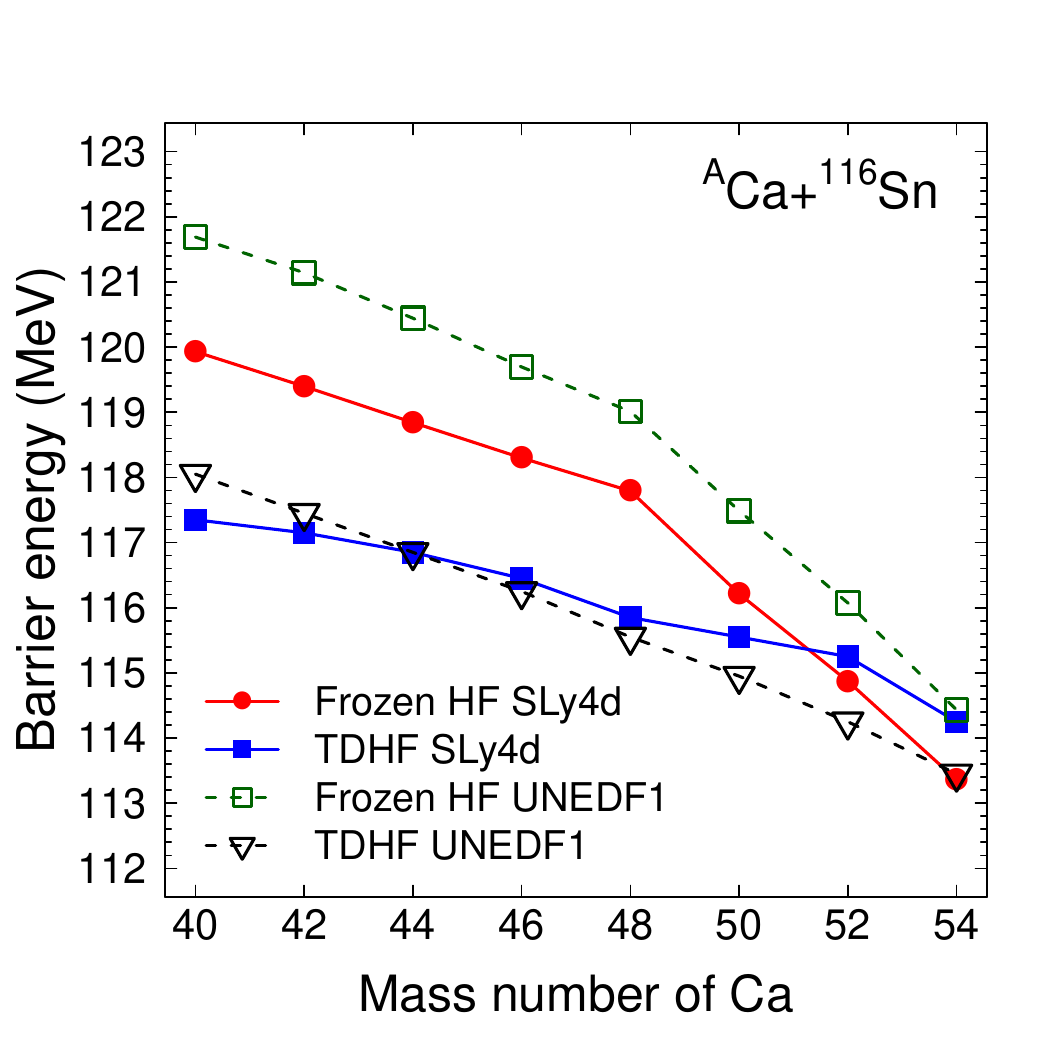}
  \caption{ (Color online) Bare potential barrier energies from the frozen HF method and TDHF 
    fusion thresholds for $^{A}$Ca$+^{116}$Sn are plotted with respect to the 
    calcium mass number for the \slyd (solid lines) 
    and \unedf (dashed lines) parametrisations.}
  \label{fig:hfall}
\end{figure}

The TDHF fusion thresholds are plotted with the corresponding frozen HF fusion
barriers in Fig.  \ref{fig:hfall} for the \slyd and \unedf parametrisations.  In
both parametrisations, we can see that for the reactions with $^{40-50}$Ca
projectiles, including dynamics has lowered the fusion barrier overall.  For all
systems it is noticeable that dynamic effects override the static effects seen
in the frozen HF barriers as there is no longer a change of slope in $V_B(A)$
near $^{48}$Ca.  Importantly, this means that the sub-barrier fusion enhancement
expected from the neutron-skin in a static picture is in fact not present when
dynamic processes are included.

For \slyd, the fusion barrier is actually increased by the dynamics for
$^{52,54}$Ca projectiles.  This is surprising, as it is expected that dynamics
should in general lower the fusion barrier \cite{hag12}.  For \unedf, the fusion
barrier has been lowered in all cases, however with a much smaller magnitude for
the most exotic calcium isotopes.
Therefore, both interactions predict that for the most neutron rich isotopes 
a dynamical mechanism occurs which counterbalances the usual lowering 
of fusion thresholds due to couplings. 

TDHF calculations intrinsically incorporate a wide variety of dynamical effects,
such as couplings to vibration and transfer channels.  It is therefore desirable
to investigate how individual dynamical effects modify the fusion barrier.  This
question is addressed in the next two sections.

\section{Vibrational couplings}\label{sec:cc}

Our aim in this section is to investigate the effect of vibrational couplings on
the fusion barrier.  The TDHF approach includes all types of dynamical
couplings, but only at the mean-field level, and without the possibility to
disentangle each contribution in a straightforward way.  Therefore, we use a
method based on a comparison between standard TDHF and coupled-channel
calculations with frozen HF and explicitly including particular dynamical modes
for example vibrations, developed in Ref.~\cite{sim13c}, to
investigate the importance of low-lying vibrations on the fusion barrier. 
The inputs to enable couplings to the collective states are obtained from TDHF.

\subsection{Nuclear vibrations}
\label{subsec:vib}

Coupled channel calculations require knowledge of the energy of the collective
states as well as their transition strengths.  Both quantities can be obtained
from TDHF calculations of a single nucleus \cite{blo79}.  This method has been
widely applied to study giant-resonances
\cite{uma05,mar05,nak05,fra12,sca13b,sca14b}, but more rarely to low-lying
vibrations \cite{sim13b,sim13c,sca13b}.  Although TDHF calculations can be used
to study non-linear vibrations \cite{rei07,sim09}, the extraction of the
transition strength relies on the linear regime, in which case it is equivalent
to the random phase approximation (RPA).  Note that TDHF in coordinate space
allows for particle evaporation \cite{cho87,ave13} and thus the escape width is
included.  The spreading width, however, involves two-body mechanisms not
accounted for in TDHF.  As before, only the initial static pairing correlations
are included.


A basic outline of linear response theory follows.  Let us consider a multipole
moment $\hat{Q}_\lambda$ of multipolarity $\lambda$.  The transition amplitude
between the ground state $|0\rangle$ with energy $E_0$ and the excited state
$|\nu\rangle$ with energy $E_\nu$ is defined as
$q_\nu=\langle\nu|\hat{Q}_\lambda|0\rangle$.  In order to calculate the
transition probability $|q_\nu|^2$, a small excitation is applied on the nucleus
at the initial time with a boost
of the form
\begin{align}
|\Psi(0)\rangle=e^{-i\varepsilon\hat{Q}_\lambda}|0\rangle,
\end{align}
where $\varepsilon$ is the boost velocity and quantifies the intensity of the
excitation.  The boost induces an oscillation of the multipole moment
expectation value which is given by
\begin{align}
   \langle\hat{Q}_\lambda\rangle(t) = -2\veps\sum_\nu |q_\nu|^2\sin[(E_\nu-E_0)t/\hbar]+O(\varepsilon).
  \label{eq:moment}
\end{align}
The linear regime is obtained by choosing $\varepsilon$ small enough so that
$\langle\hat{Q}_\lambda\rangle$ is linearly proportional to $\varepsilon$.  
The strength function is
then computed from a sine Fourier transform of
$\langle\hat{Q}_\lambda\rangle(t)$
\begin{eqnarray}
Q_\lambda(E)&=& \lim_{\varepsilon\rightarrow0}\frac{-1}{\pi\hbar\varepsilon} \int_0^\infty dt \,\,\langle\hat{Q}_\lambda\rangle(t) \sin (Et/\hbar) \label{eq:strength1} \\
&\simeq& \sum_\nu|q_\nu|^2\delta[E-(E_\nu-E_0)].\label{eq:strength2}
\end{eqnarray}
In practice, $\langle\hat{Q}_\lambda\rangle(t)$ is only computed over a finite
time.  To avoid spurious oscillations in the strength function,
$\langle\hat{Q}_\lambda\rangle(t)$ is filtered in the time domain by multiplying
it by a scaled half-Gaussian damping function reaching zero at the end of the calculation
\cite{mar05}.  This damping function adds only a small width to the peaks in the
strength function.

We focus on quadrupole and octupole vibrations which can strongly couple to the
relative motion.  The quadrupole and octupole operators are defined as
\begin{align*}
\hat{Q}_{2} &= \sqrt{\frac{5}{16\pi}} \sum_{i=1}^A
    \left(2\hat{x}^2-\hat{y}^2-\hat{z}^2\right) \\
\hat{Q}_{3} &= \sqrt{\frac{7}{16\pi}}\sum_{i=1}^A
  \left[2\hat{x}^3-3\hat{x}\left(\hat{y}^2+\hat{z}^2\right)\right],
\end{align*}
respectively.  The operators $\hat{x}$, $\hat{y}$ and $\hat{z}$ are
single-particle observables and the sums run over each nucleon.

\begin{figure}[h]
  \graphicspath{ {./pix/} }
    \includegraphics[width=8cm,page=1]{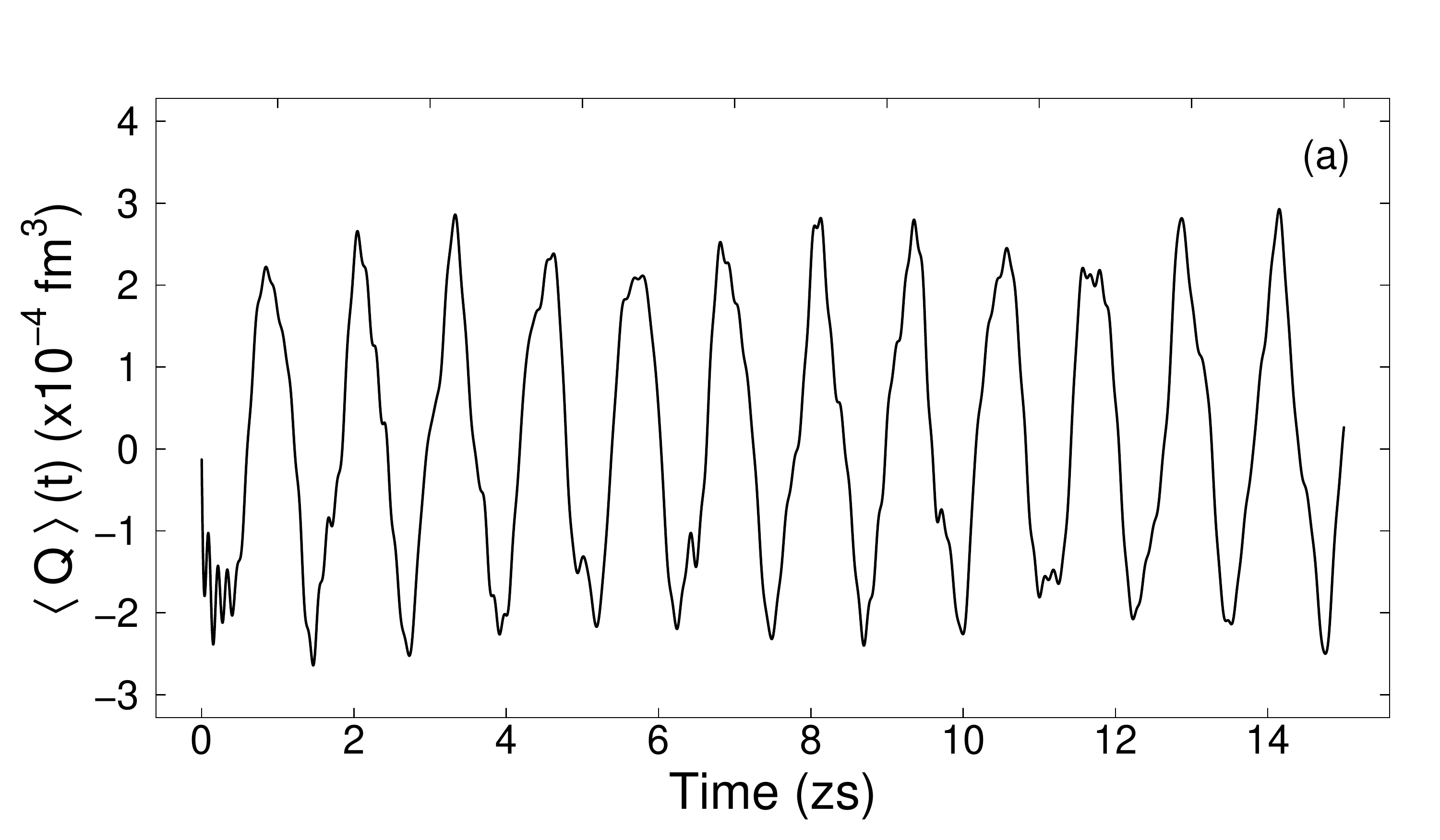}
    \includegraphics[width=8cm,page=3]{40Ca_octu_1e-8_n.pdf}
  \caption{(a) Time evolution of the octupole moment induced by an octupole
    boost on $^{40}$Ca in the linear regime.  (b) Associated strength function
    computed from Eq.~(\ref{eq:strength1}).}
  \label{fig:octu}
\end{figure}

All TDHF vibration calculations were performed using the \slyd
interaction in the same box size as for HF ground states in section~\ref{sec:static}
but with one plane of symmetry $z=0$, and
for a total time of 15 zs.  An example of evolution of the octupole
moment following an octupole boost on $^{40}$Ca is presented in
Fig.~\ref{fig:octu}(a).  A strong oscillation is observed, producing
an intense peak at 3.44~MeV in the strength function plotted in
Fig.~\ref{fig:octu}(b).  This peak is associated to the $3^-_1$ first
phonon of the low-lying collective octupole mode.  The transition
probability $|q_\nu|^2$ is obtained from the area of the peak and
transformed into a deformation parameter according to
\begin{align}
  \beta_\lambda = \frac{4\pi|q_\nu|}{3AR_0^\lambda} .
  \label{beta}
\end{align}
The radius $R_0$ is taken as the radius of the isodensity surface at half the
saturation density $\rho_0/2=0.08$~fm$^{-3}$.

\begin{table}[h]
  \centering
  \begin{tabular}{c c c c c}
    \hline\hline
    $A$ (Ca) & \multicolumn{2}{c}{\hspace{0.3cm}TDHF} \hspace{0.5cm}& \multicolumn{2}{c}{\hspace{0.5cm}Experiment}\hspace{0.5cm} \\ \hline
      & $E_3$ & $\beta_3$ & $E_3$ & $\beta_3$ \\
    40 & 3.44 & 0.224 & 3.737 & $0.30-0.41$ \\
    42 & 4.14 & 0.195 & 3.447 & 0.26 \\
    44 & 4.68 & 0.165 & 3.308 & $0.23-0.26$ \\
    46 & 5.14 & 0.141 & 3.614 & 0.16 \\
    48 & 5.52 & 0.109 & 4.507 & $0.17-0.25$ \\
    50 & 4.62 & 0.168 & - & - \\
    52 & 3.48 & 0.221 & -  & - \\
    54 & 2.92 & 0.226 & -  & - \\
    \hline\hline
  \end{tabular}
  \caption{TDHF and experimental \cite{kib02} excitation energies and
    deformation parameters for $3_1^-$ states in calcium isotopes. }
  \label{tab:oct}
\end{table}

The low-lying octupole phonon energies and their associated
deformation parameters computed with TDHF are reported in
Table~\ref{tab:oct} alongside experimental values.  The purpose of
this comparison is not to achieve the best possible description of
vibrational modes (we leave that for a future work), but rather to
verify what types of vibrational couplings are included in the TDHF
dynamics.  The TDHF results in Table~\ref{tab:oct} can be interpreted
in a simple spherical shell model picture.  The $3^-_1$ state in
$^{40}$Ca is a coherent sum of one-particle one-hole excitations
across the $N=Z=20$ magic shell gap.  Adding neutrons to the
$1f_{7/2}$ shell blocks neutrons excitations across the $N=20$ magic
gap, hindering collectivity.  This is seen as a reduction of $\beta_3$
going from $A=40$ to $48$ and results in an increase of $E_3$ due to
less residual interaction.  There is also a large energy gap between
$1f_{7/2}$ and the next positive parity level ($1g_{9/2}$).  Having
filled the $1f_{7/2}$ shell, collectivity then increases due to
increased excitations between the $fp$-shell and $1g_{9/2}$ for
$A>48$.

The above observations are in qualitative agreement with the
experimental data for doubly magic nuclei ($^{40,48}$Ca).  The
situation is more complicated for mid-shell nuclei which could be
affected by pairing correlations neglected in the RPA residual
interaction.  However, the energies agree to better than $30\%$ and
the deformation parameters, while underestimated in TDHF, are of the
same order of magnitude.  This is sufficient for the purpose of
investigating the impact of low-lying octupole vibrations in fusion.


Similar calculations have been performed for quadrupole vibrations.  No
low-lying collective oscillations are found in $^{40}$Ca as $2^+$ one-particle
one-hole states require excitations across two magic gaps (magic numbers 20 and 28), 
therefore contributing only to the giant quadrupole resonance.  Neutrons in the partially
filled $1f_{7/2}$ shell can produce low-lying $2^+$ states by breaking a pair
and coupling within the same shell.  Experimentally, this leads to $2^+_1$
states at $E_2\simeq1-1.5$ MeV in $^{42,44,46}$Ca isotopes.  However, these
states have zero excitation energy in TDHF as pairing is neglected.  The first
$2^+$ excitations in the strength functions calculated with TDHF for $40<A\le48$
calcium isotopes are then obtained by promoting $1f_{7/2}$ neutrons across the
$N=28$ gap.  For the $48<A\le54$ calcium isotopes also studied here, low-lying
$2^+$ states can be formed by one-particle one-hole excitations in the $fp$
shell.  Overall, we found that these low-lying quadrupole vibrations, as
calculated in TDHF, are globally much less collective than the octupole
modes. The largest quadrupole deformation parameter is found in the $^{48}$Ca
$2^+_1$ state at $E_2=3.828$~MeV with $\beta_2=0.078$, in good agreement with
experiment ($E_2=3.832$~MeV and $\beta_2=0.101\pm0.017$ \cite{ram87}).  We have
checked that, in TDHF, the main contribution of coupling effects on the barrier
comes from the octupole states and we therefore only include $3^-_1$ states in
the following coupled-channel calculations.


\subsection{Coupled-Channels calculations}
\label{subsec:cc}

Coupled channel (CC) calculations of $^{A}$Ca$+^{116}$Sn have been performed with the
\verb+CCFULL+ code \cite{hag99}.  The nuclear potential is taken to be in the
Woods--Saxon form,
\begin{align*}
V_N(r) = \frac{-V_0}{1+\exp{\left[(r-R)/a\right]}},
\end{align*}
with the usual three parameters of potential depth $V_0$, diffuseness $a$ and
radius $R$.  In all calculations these parameters were taken from fitting the
Woods--Saxon potential to the frozen HF bare potential obtained with the \slyd
interaction to reproduce the barrier energy within a 1~keV error.  The values of
the Woods-Saxon parameters are given in Table \ref{tab:CCWS}.  The energy and
deformation parameter of the $3^-_1$ state were taken from the TDHF results in
Table \ref{tab:oct}.
No transfer coupling was included in any of the CC calculations.

The barrier distribution $D(E)$ has been calculated from the \verb+CCFULL+
fusion cross-section using
\cite{row91}
\begin{align*}
D(E) = \frac{d^2(E\sigma)}{dE^2}.
\end{align*}
The barrier distribution is positive for energies ranging from 0 MeV up until
some particular energy $E'$, above for which it becomes negative \cite{das98}.  The
average fusion barrier is then calculated using the centroid of $D(E)$ with the
upper integration limit of $E$ being $E'$,
\begin{align}
  V_B = \frac{\int_0^{E'} E D(E) dE}{\int_0^{E'} D(E) dE}.
  \label{eq:VbD}
\end{align}



\begin{table}[h]
  \centering
  \begin{tabular}{c c c c}
    \hline\hline
     $A$ (Ca) & $V_0 (MeV)$ & $R (fm)$ & $a (fm)$\\ \hline
    40 & 76.433 & 1.199 & 0.611 \\ 
    42 & 72.773 & 1.202 & 0.604 \\
    44 & 75.171 & 1.198 & 0.603 \\
    46 & 73.054 & 1.199 & 0.600 \\
    48 & 75.254 & 1.195 & 0.599 \\
    50 & 90.939 & 1.179 & 0.641 \\
    52 & 102.237 & 1.170 & 0.667 \\
    54 & 125.215 & 1.153 & 0.701 \\
    \hline\hline
  \end{tabular}
  \caption{Woods--Saxon fit parameters for frozen HF nucleus-nucleus potential
    of $^A$Ca+$^{116}$Sn reactions with the \slyd interaction. }
  \label{tab:CCWS}
\end{table}

\begin{figure}[h]
  \centering
  \graphicspath{ {./pix/} }
    \includegraphics[width=8cm, page=2]{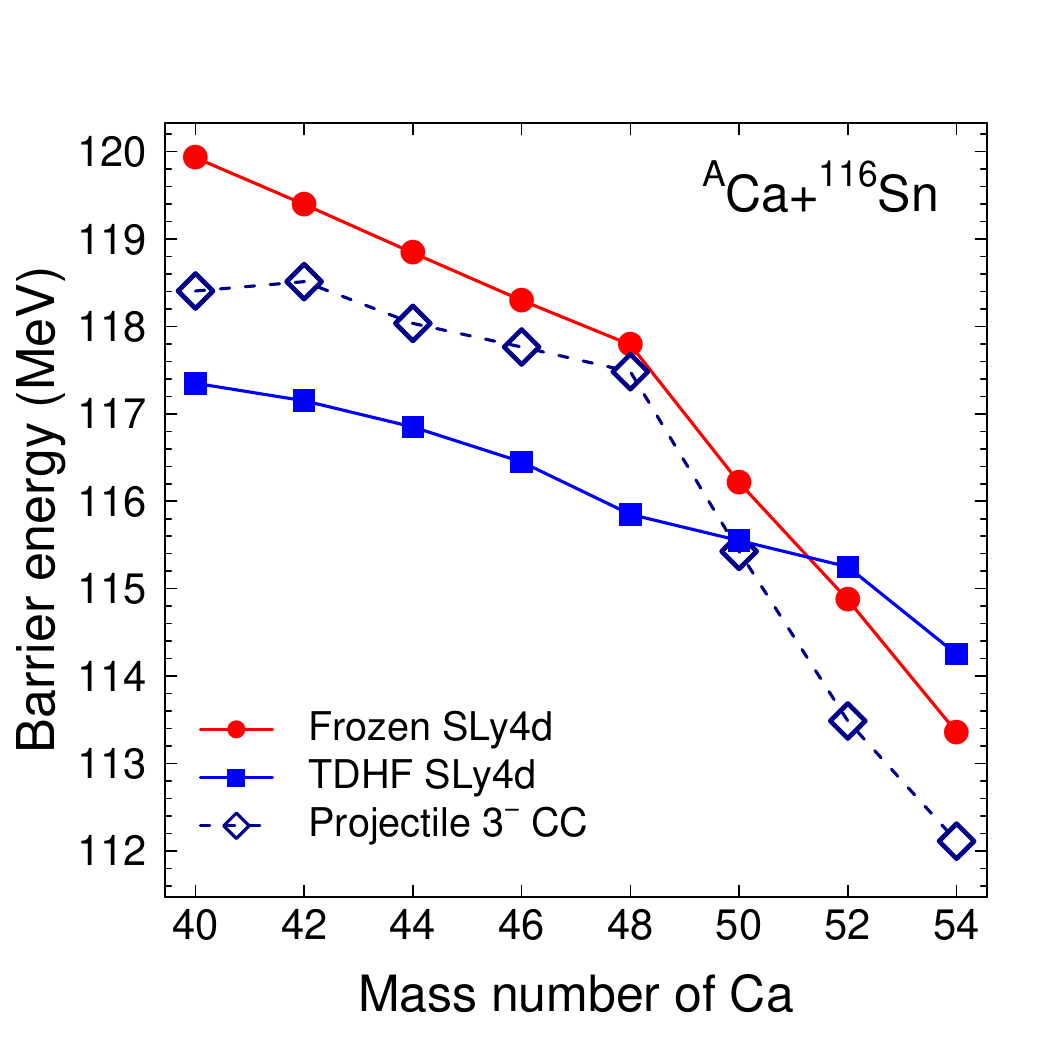}
  \caption{(Color online) Frozen HF (cirlces) barriers in $^{A}$Ca+$^{116}$Sn are compared with
    TDHF fusion thresholds (squares) and coupled-channels average barriers with
    couplings to the $3^-_1$ state in calcium isotopes (diamonds).
   }
  \label{fig:ccfulls}
\end{figure}

We have checked that, when no couplings are included in the \verb+CCFULL+
calculations, the resulting centroid of the barrier distribution $D(E)$ matches
the frozen-HF barrier.  Inclusion of coupling to the first octupole phonon in
the calcium isotopes systematically reduces the centroid of $D(E)$ by up to
$\sim1.5$~MeV, as shown in Fig.~\ref{fig:ccfulls}.

The impact of coupling to low-lying quadrupole phonons in calcium isotopes has
also been studied using experimental data on the $2^+_1$ state (when available).
Although this coupling may modify the shape of the barrier distribution, it
barely affects its centroid.  Higher energy states, such as giant quadrupole
resonances have only a small impact on the barrier \cite{hag97,sim13c}.
Coupling to the $3_1^-$ state in $^{116}$Sn further reduces the barrier
for the systems $^{42-54}$Ca+$^{116}$Sn by less than $0.2$~MeV.




The CC calculations confirm that the vibrational states
included in the TDHF calculations lower the fusion threshold.
This helps explain the lowering of the barrier due to dynamical
effects as observed in Fig.~\ref{fig:hfall} for reactions with the
non-exotic calcium isotopes.  However, vibrational couplings cannot
explain why the fusion thresholds is actually higher than the frozen
HF barrier barrier (with \slyd) for the most exotic nuclei.

\section{Role of transfer}\label{sec:transfer}

While vibrational couplings usually lower the barrier, the effect of transfer
channels is less clear despite several experimental investigations
\cite{Beckerman80,Stefanini86,Timmers97,eve11,bou14,jia14,Jiang15,trz16}.  One
problem is the difficulty to incorporate transfer channels in coupled-channel
calculations \cite{tho85,pol13,kar15,sca15b}.  Alternatively, microscopic
approaches can also be used to study transfer reaction mechanisms in heavy-ion
collisions \cite{koo77,sim10b,ayi15}.  Here, our study of transfer channels is
motivated by the observation of an increase of the fusion barrier in
$^{52,54}$Ca+$^{116}$Sn in TDHF calculations and which cannot be explained by
vibrational couplings (see Fig.~\ref{fig:ccfulls}).


Transfer probabilities are computed using the particle number projection
technique developed in Ref.~\cite{sim10b} for systems without pairing, and
extended in Ref.~\cite{sca13a} for superfluid systems.  This method has been
used to study multi-nucleon transfer reactions
\cite{sim10b,sca13a,sek13,sca15c,son15,bou16} and fission \cite{sim14,sca15a}.
Here we use it to determine proton transfer probabilities in
$^A$Ca$+^{116}$Sn. As the fragments both have magic proton numbers, the
resulting proton transfer probabilities are not affected by pairing correlations
so we use the simple projection technique \cite{sim10b}.

\begin{figure}[h]
  \centering
  \graphicspath{ {./pix/} }
  \includegraphics[scale=0.4, clip=true, trim= 0 0 0 17mm]{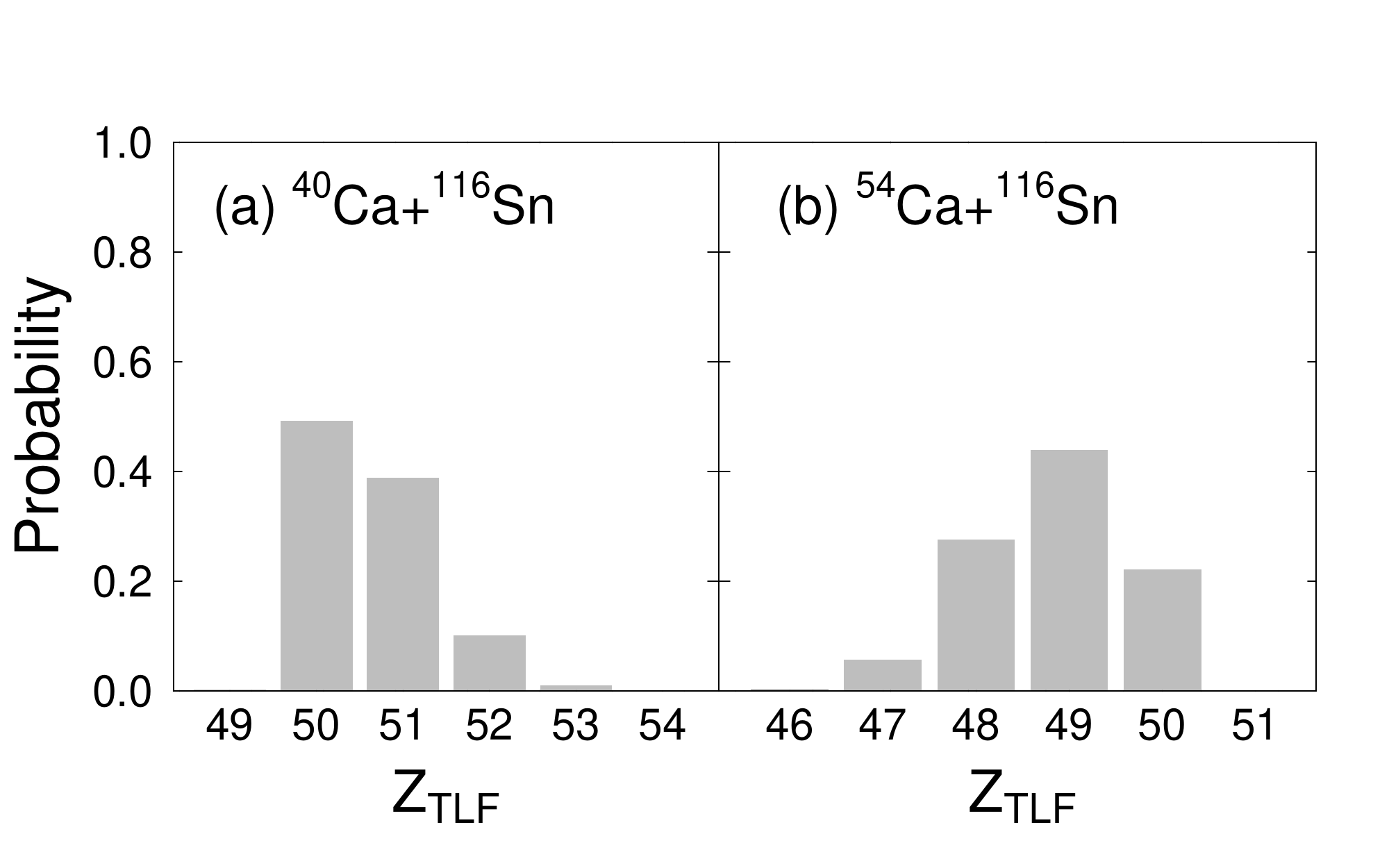}
  \caption{Proton number probability distributions in the outgoing TLF
     in $^{40}$Ca$+^{116}$Sn (a) and $^{54}$Ca$+^{116}$Sn (b) central
    collisions at an energy of $99.9\%$ of the TDHF fusion threshold.}
  \label{fig:transf2}
\end{figure}

All calculations were again made using the \slyd interaction.  The
probability distribution of the final proton number in the target-like
fragment (TLF) is shown in Fig.~\ref{fig:transf2}(a) for
$^{40}$Ca$+^{116}$Sn and in Fig.~\ref{fig:transf2}(b) for
$^{54}$Ca$+^{116}$Sn at an energy of $99.9\%$ of the TDHF fusion
threshold.  As seen in Fig.~\ref{fig:transf2}(a), protons are
transferred from the light fragment to the TLF in
$^{40}$Ca$+^{116}$Sn with a probability of $\sim50\%$.  Conversely
Fig.~\ref{fig:transf2}(b) shows that $^{116}$Sn loses protons in
$^{54}$Ca$+^{116}$Sn, with only $\sim20\%$ chance to find a tin
fragment in the exit channel.

\begin{figure}[h]
  \centering
  \graphicspath{ {./pix/} }
  \includegraphics[page=2,scale=0.4, clip=true, trim = 0 0 0 2cm]{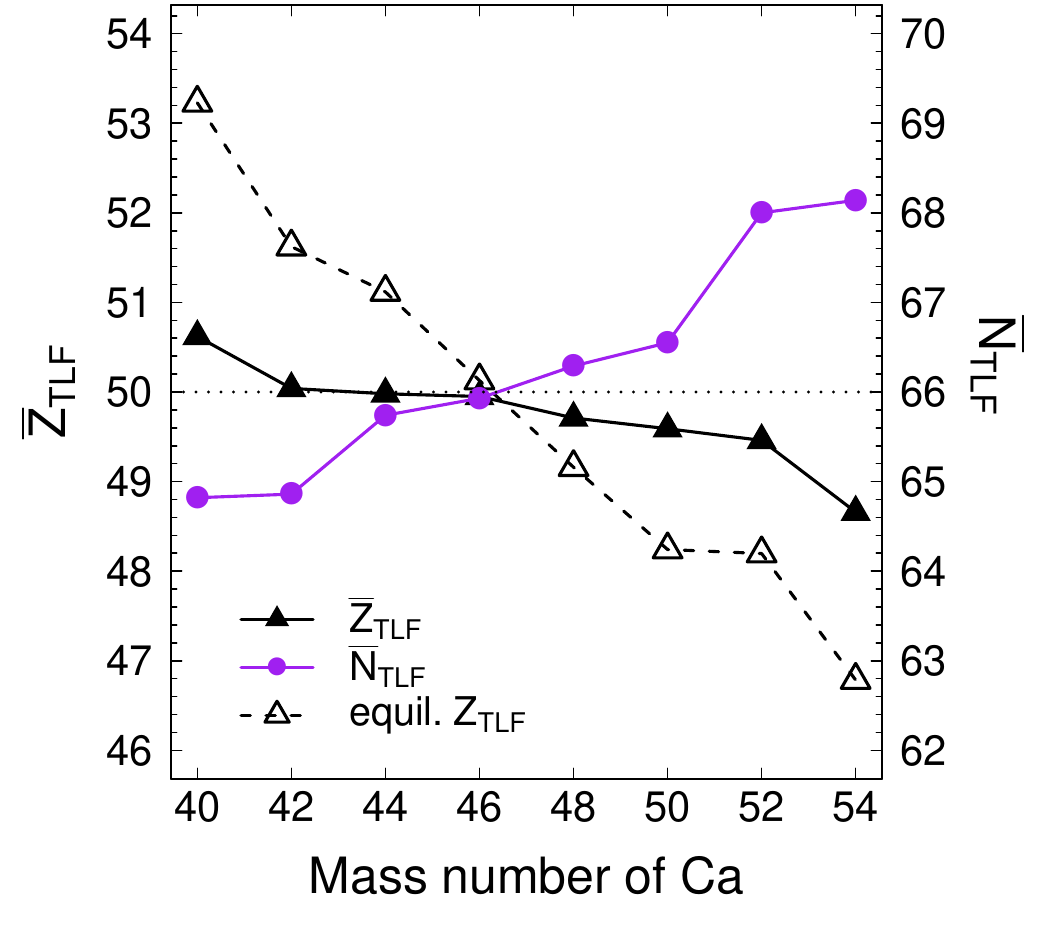}
  \caption{(Color online) The average proton (left axis) and neutron number (right
    axis) of the TLF.  The dashed line and open
    triangles show the anticipated $\overline{Z}_\text{TLF}$ value assuming the
    system is fully equilibriated with the TLF having
    $N_\text{TLF}=\overline{N}_\text{TLF}$.  The original target $Z$ and $N$ (horizontal
    dotted line) are also shown.
    }
  \label{fig:transf1}
\end{figure}

A signature of transfer reactions can also be obtained from the average of the
number of nucleons in the final fragments, which is simply determined by
integrating the proton and neutron densities around one fragment in the exit
channel.  Figure~\ref{fig:transf1}(a) shows both the average proton
($\overline{Z}_\text{TLF}$) and neutron ($\overline{N}_\text{TLF}$) numbers in the TLF. 
 We observe that $\overline{Z}_\text{TLF}$ decreases while $\overline{N}_\text{TLF}$
increases with increasing calcium mass number, confirming the results
in Fig.~\ref{fig:transf2}.


The direction of the transfer is determined by a charge equilibration
process where the initial neutron to proton ratio $N/Z$ asymmetry
between the fragments is reduced after contact.  This is a
manifestation of positive $Q$-values for transfer reactions induced by
the symmetry energy, studied in detail with TDHF in transfer reactions
\cite{sim12a,sek13}.  The dashed line in Fig \ref{fig:transf1} shows
the equilibriated numbers of protons for the given $\overline{N}$, assuming
that the projectile and target both have the $N/Z$ of the compound
system.  This line indicates that, at this collision energy, the
transfer reactions do no fully equilibriate the reactants.  However, it
should be regarded as an upper limit since the equilibration acts to
increase binding rather than truly equilibriate the neutron to proton
ratio.


The present calculations indicate that neutron transfer is stronger
than proton transfer in this process, resulting in net mass transfer
to the light calcium isotopes and from the heavier isotopes.  As shown
in Fig. \ref{fig:radii}, the rms radii of the neutrons in the calcium
isotopes are generally larger than those for the protons, making them
more accessible for transfer.  The influence of neutron transfers on
fusion is not fully understood.

When the proton transfer to calcium occurs the charge product of the
fragments increases which in turn
increases the Coulomb repulsion and thus the fusion barrier.  
This suggests a possible mechanism for the increase of the fusion threshold due
to dynamical effects in $^{52,54}$Ca+$^{116}$Sn.

An alternative explanation would be that dissipation of the initial
kinetic energy is faster (meaning it occurs at larger distances) with
calcium isotopes with $A>48$ due to a larger level density near the
Fermi level and weak neutron binding.  Further studies are required to
better understand the role of transfer and dissipation in the
dynamical effects on the fusion barrier.  For instance, a simple proxy
to the dissipation can be obtained in TDHF from the total kinetic
energy loss \cite{obe14} and from the number of emitted nucleons
\cite{ked10}.  More advanced techniques to extract the energy
dissipated into excitation energies include a macroscopic reduction
procedure \cite{was09c}, the density-constrained TDHF approach
\cite{uma09a}, and a more general application of the particle number
projection technique \cite{sek14}.

\section{Conclusion}

A systematic study of fusion barriers in reactions between a stable target
($^{116}$Sn) and a chain of calcium projectiles ranging from stable to unstable
neutron rich isotopes has been performed using microscopic approaches based on
the Hartree-Fock mean-field approximation.

The bare potential barriers obtained assuming frozen ground-state densities
decrease with the calcium mass number.  The results also show that the
development of a neutron skin for calcium isotopes heavier than $^{48}$Ca
further decreases the bare barrier.

However, this static effect on the bare barrier disappears when
dynamic effects are taken into account via the TDHF approach.  The
inclusion of dynamical effects globally lowers the fusion threshold
except for reactions with the most exotic calcium isotopes studied
here ($^{52,54}$Ca).  Depending on the choice of the Skyrme
interaction, the fusion threshold can even become higher than the bare
barrier for these exotic projectiles.

Coupled channels calculations have been performed to understand the
contribution of couplings to low-lying vibrations to the dynamical
modification of the barrier.  The results show that vibrational
couplings systematically lower the average barrier and are thus not
responsible for the increase of barrier energy which will
 hinder fusion with exotic calcium projectiles.

The importance of transfer channels near the barrier has also been
investigated with TDHF calculations for these systems.  The results,
which can be interpreted in terms of a simple charge equilibration
process, suggest that the Coulomb repulsion is increased due to charge
transfer in $^{52,54}$Ca$+^{116}$Sn.  This mechanism provides a
possible explanation for the fusion hindrance in these systems.  An
increase of dissipation associated with a weakly bound collision
partner could also provide a mechanism for fusion hindrance.

More work is needed to get a deeper insight into the role of transfer
and dissipation mechanisms in microscopic dynamics.  It is also
desirable to include dynamic pairing correlations as this would affect
inelastic excitations and multinucleon transfer probabilities.
Various Skyrme parametrisations and effective interactions should also
be tested as they may lead to different predictions for reactions with
exotic beams.  For instance, reactions with the Gogny interaction have
recently been performed \cite{sca16}.  The use of effective
interactions from the quark-meson coupling model has also shown
promising results in static HF calculations of nuclear structure
\cite{sto16} and could be implemented in TDHF codes.

\acknowledgements The authors are grateful to D. J. Hinde, M. Dasgupta, and
A. S. Umar for stimulating discussions during this work.  This research was
undertaken with the assistance of resources from the National Computational
Infrastructure (NCI), which is supported by the Australian Government.  This
research was supported under Australian Research Council's Future Fellowship
(project number FT120100760), Discovery Projects (project number DP140101337),
and Laureate Fellowship (project number FL110100098) funding schemes.

\pagebreak

\bibliography{ref}
\end{document}